\documentclass{elsart}

\usepackage{amsfonts,amssymb}
\usepackage{graphicx}  

\begin{document}
\begin{frontmatter}
\title{Temperature-dependent bandstructure of bulk EuO}
\author{R.~Schiller and W.~Nolting}
\address{Humboldt-Universit\"at zu Berlin, Institut f\"ur Physik,
            Invalidenstra{\ss}e 110, D-10115 Berlin, Germany}
\begin{abstract} 
We present calculations for the temperature-dependent electronic
structure of bulk ferromagnetic EuO based on a parametrization of the
{\em d-f} model Hamiltonian using results of first-principles TB-LMTO
band structure calculations. The presented method avoids the problem of
double-counting of relevant interactions and takes into account the
symmetry of the atomic orbitals. It enables us to determine the
temperature-dependent band structure of EuO over the entire temperature
range. 
\end{abstract}
\end{frontmatter}
The model of choice for describing the electronic and magnetic
properties of magnetic rare-earth systems is the {\em s-f} 
(or {\em s-d}) model \cite{Nol79a}. The model describes the exchange
coupling of itinerant band electrons to localized magnetic moments.
It is applicable to the magnetic semiconductors as the europium
chalcogenides and to the metallic elemental rare earths Gd, Tb, and Dy.
Many characteristics of these materials can be explained by a
correlation between the localized magnetic {\em 4f}-states and extended
conduction band states, mainly {\em 5d-6s}. Within the {\em s-f} model,
the correlation is represented by an intraatomic exchange interaction.

There have been earlier works on the temperature-dependent band
structure of bulk EuO \cite{NBB87a,NBB87b}. The presented approach
employs a decomposition of the relevant Eu-{\em 5d} bands of EuO into
five consecutive bands. For each of these bands, a single-band {\em s-f}
model calculation has been performed. By employing this splitting
procedure, the method employed by the authors \cite{NBB87a,NBB87b}
disregards the full multi-band structure of the conduction bands in EuO
and leads to an overestimation of relevant correlation effects.

In the following, we will introduce a multi-band {\em d-f}
model and use it to calculate the temperature-dependent bandstructure of
EuO. The employed method combines a reliable many-body treatment of the
{\em d-f} exchange model with a self-consistent band-structure
calculation. In particular, the full multi-band aspect of the problem is
treated in a reliable and symmetry-conserving mode.

\section{Theory}
The conventional or single-band {\em s-f} model describes the intratomic
exchange interaction between an {\em s}-like conduction band and system
of localized magnetic moments, which are due to the partially filled
{\em 4f}-shells. For a real material, like EuO, the conduction band
usually consists of multiple subbands, which can be distinguished by
their magnetic quantum number. In the following we will introduce the
model for such a local-moment system with multiple conduction bands. The
respective model will be referred to as the (multi-band) {\em d-f} model
as opposed to the (single-band) {\em s-f} model.

The {\em d-f} model Hamiltonian consists of three parts:
\begin{equation}
  \label{ham}
  H = H_d + H_f + H_{df}.
\end{equation}
For a real system like EuO, the kinetic energy of the conduction
electrons reads: 
\begin{equation}
  \label{h_d}
  H_d = \sum_{ijmm'} T^{mm'}_{ij} c^+_{im\sigma} c_{jm'\sigma}.
\end{equation}
Here, $c^+_{im\sigma}$ and $c_{im\sigma}$ are, respectively, the
creation and annihilation operators of an electron from the $m$-th
subband at the lattice site ${\bf R}_i$. $T^{mm'}_{ij}$ are the hopping
integrals. 

Each lattice site ${\bf R}_i$ is occupied by a localized magnetic
moment, represented by a spin operator ${\bf S}_i$. 
These localized moments
are exchange coupled expressed by the Heisenberg Hamiltonian:
\begin{equation}
  \label{h_f}
  H_f = \sum_{ij} J_{ij} 
  {\bf S}_i \cdot {\bf S}_j,
\end{equation}
where $J_{ij}$ are the Heisenberg exchange integrals.

In addition to the contribution of the itinerant-electron system and the
contribution of the localized {\em f}-moments we have a third contribution to
account for an intraatomic interaction between the conduction electrons and the
localized {\em f}-spins. The form of that contribution can be derived
from the general form of the Coulomb interaction between electrons at a
single lattice sites. If the electron scattering processes caused by the
Coulomb interaction are restricted to two involved subbands $L$ and
$L'$, one gets for the Coulomb interaction:
\begin{eqnarray}
  \label{coulomb}
  H_I = \textstyle\frac{1}{2}\displaystyle
  \sum_{LL'\sigma\sigma'} \big[ U_{LL'} c^+_{L\sigma} c^+_{L'\sigma'}
  c_{L'\sigma'} c_{L\sigma}
  & + & J_{LL'} c^+_{L\sigma} c^+_{L'\sigma'}
  c_{L\sigma'} c_{L'\sigma}\nonumber\\
  & + & J^*_{LL'} c^+_{L\sigma} c^+_{L\sigma'}
  c_{L'\sigma'} c_{L'\sigma} \big].
\end{eqnarray}
We can split the Coulomb interactions into three different parts,
depending on whether both $L$ and $L'$ are conduction bands, 
$H_I^{CC}$, both belong to localized bands, $H_I^{LL}$, or
there is an interaction between localized and conduction bands involved,
$H_I^{CL}$,
\begin{equation}
  \label{coulomb_split}
  H_I = H_I^{CC} + H_I^{LL} + H_I^{CL}.
\end{equation}
The first of the three parts disappears, since for the case of a
semiconductor the conduction bands are unoccupied. The second term is
already contained in the description of the localized moments via the
Heisenberg model in Eq.~(\ref{h_f}). 
The third term is the only one which contains an
interaction between the localized bands and the conduction
bands. Assigning the indices $m$ and $f$ to itinerant and localized
bands, respectively, it can be written in the form: 
\begin{eqnarray}
  \label{coulomb_cl}
    H_I^{CL} = \sum_{mf\sigma\sigma'}
    \Big[ & & U_{mf} c^+_{m\sigma} c^+_{f\sigma'} c_{f\sigma'} c_{m\sigma}
    + J_{mf} c^+_{m\sigma} c^+_{f\sigma'} c_{m\sigma'}
    c_{f\sigma}\nonumber\\ & & + \textstyle \frac{1}{2} 
    \displaystyle \big( J^*_{mf} c^+_{m\sigma} c^+_{m\sigma'} 
    c_{f\sigma'} c_{f\sigma} + J^*_{fm} 
    c^+_{f\sigma} c^+_{f\sigma'} c_{m\sigma'} c_{m\sigma} \big) \Big]. 
\end{eqnarray}
The last two terms of the above equation vanish for the special case of
${\rm EuO}$, due to the two fact that in ${\rm Eu}$ the localized 
{\em 4f} shell has its maximum spin of 
$S=7/2$ resulting in no double occupancy of the different subbands of
the {\em 4f} shell. Introducing the Pauli spin operators
$\sigma_L$ one arrives at
\begin{equation}
  \label{coulomb_fin}
  H_I^{CL} = -2\sum_{mf} J_{mf} \, \sigma_m \cdot
  \sigma_f.
\end{equation}  
By defining the spin operator of the local moment by 
${\bf S}=\hslash\sum_f \sigma_f$ and by assuming the
inter-band exchange to be independent on the band indices $m$ and $f$,
$J_{mf}\equiv J/2$, the multi-band {\em d-f} interaction which is the
sum of Eq.~(\ref{coulomb_fin}) over all the different lattice sites
reads:
\begin{equation}
  \label{h_df_multi}
  H_{df} = - \frac{J}{\hslash} \sum_{im} 
  \sigma_{im} \cdot {\bf S}_{i}.
\end{equation}
where $J$ is the intraatomic {\em d-f} exchange interaction.
Using the second-quantized form of $\sigma_{i}$ and the abbreviations
\begin{equation}
\label{ladderop}
S^{\sigma}_{j} = S^x_{j} + {\rm i} z_{\sigma} S^y_{j};
\quad z_{\uparrow(\downarrow)} = \pm 1,
\end{equation}
the {\em d-f} Hamiltonian can be written as
\begin{equation}
  \label{h_df}
  H_{df} =  - \frac{J}{2} \sum_{im\sigma} \left( z_{\sigma}
    S^z_{i} n_{im\sigma} + S^{\sigma}_i
    c^+_{im-\sigma} c_{im\sigma} \right).
\end{equation}
The most decisive part of the {\em d-f} Hamiltonian (\ref{h_df}) is the 
second term,
which describes spin exchange processes between the conduction electrons
(\ref{h_d}) and the localized moments (\ref{h_f}).

The many-body problem that arrises with the Hamiltonian~(\ref{ham}) is
far from being trivial and a full solution is lacking. In previous
papers we have presented an approximate treatment for single-band 
{\em s-f} model in a local-moment system with film geometry for the
special case of an empty conduction band, $n=0$ \cite{SN99b}. 
For this special case, the approach for the single-band {\em s-f} model
in a film geometry with inequivalent layers $\alpha$ can be identically
transferred to the situation of the multi-band {\em d-f} model in bulk
materials with different subbands $m$, via the index transition $\alpha
\rightarrow m$. Using this transition, in the following we want to
present the main results of the theoretical approach for the multi-band
{\em d-f} model.

Due to the empty conduction bands we are considering throughout the
whole paper, the Hamiltonian  (\ref{ham}) can be split into an
electronic part, $H_d + H_{df}$, and a magnetic part,
$H_f$, which can be solved separately \cite{SN99b}. 
For the magnetic subsystem $H_f$ we employ an approach based on
the random phase approximation (RPA) which has been described in detail
in \cite{SN99a}. As the result one gets the temperature-dependent
magnetizations of the local-moment system. 

Concerning the electronic subsystem, $H_d + H_{df}$, all physically
relevant information can be derived from the retarded single-electron
Green function 
\begin{eqnarray}
  \label{segf}
  G^{mm'}_{ij\sigma}(E) & = & \left\langle\!\left\langle c_{im\sigma} ;
      c^+_{jm'\sigma} \right\rangle\!\right\rangle_E\nonumber\\
  & = & - {\rm i} \int\limits_0^{\infty} dt {\rm e}^{-\frac{\rm
      i}{\hslash} Et} \left\langle [ c_{im\sigma}(t) ,
    c^+_{jm'\sigma}(0) ]_+ \right\rangle.
\end{eqnarray}
The Fourier transform of the single-electron Green function,
$G^{mm'}_{{\bf k}\sigma}(E)= N^{-1} \sum_{ij} {\rm e}^{{\rm i} {\bf k}
    ({\bf R}_i-{\bf R}_j)} G^{mm'}_{ij\sigma}(E)$, 
is related to the spectral density via
\begin{equation}
  \label{spdens}
  S^{mm'}_{{\bf k}\sigma}(E) = -\frac{1}{\pi} {\rm Im} 
  G^{mm'}_{{\bf k}\sigma}(E),
\end{equation}
from which the partial density of states of the $m$-th subband can be
obtained: 
\begin{equation}
  \label{pdos}
  \rho^m_{\sigma}(E) = \frac{1}{\hslash N} \sum_{\bf k} 
  S^{mm}_{{\bf k}\sigma}(E).
\end{equation}

For the solution of the many-body problem posed by the Hamiltonian 
$H_d + H_{df}$ we write down the equation of motion of the
single-electron Green function 
(\ref{segf})
\begin{equation}
\label{eom_gen}
E\,G^{mm'}_{ij\sigma} = \hslash \delta_{mm'} \delta_{ij} +
\sum_{km''} T^{mm''}_{ik} G^{m''m'}_{kj\sigma} + 
\langle\!\langle [c_{im\sigma},H_{df}]_-;
c^+_{jm'\sigma} \rangle\!\rangle_E,
\end{equation}
The formal solution of Eq.\ (\ref{eom_gen}) can be found by introducing
the self-energy $M^{mm'}_{ij\sigma}(E)$,
\begin{equation}
  \label{self_gen}
  \left\langle\!\left\langle \left[c_{im\sigma},H_{df}\right]_-;
  c^+_{jm'\sigma} \right\rangle\!\right\rangle_E =
  \sum_{km''} M^{mm''}_{ik\sigma}(E) G^{m''m'}_{kj\sigma}(E),  
\end{equation}
which contains all information about the correlations between the conduction
band and localized moments. After combining Eqs.\ (\ref{eom_gen}) and 
(\ref{self_gen}) and performing a Fourier transform we
see that the formal solution of Eq.\ (\ref{eom_gen}) is given by
\begin{equation}
  \label{form_sol}
  {\bf G}_{{\bf k}\sigma}(E) = \hslash \left(E\,{\bf I}-{\bf T}_{\bf k} -
  {\bf M}_{{\bf k}\sigma}(E) \right)^{-1},
\end{equation}
where ${\bf I}$ represents the ($M \times M$) identity matrix and where
$M$ is the number of subbands of the condcution-band system and where
the matrices ${\bf G}_{{\bf k}\sigma}(E)$, ${\bf T}_{\bf k}$, and 
${\bf M}_{{\bf k}\sigma}(E)$
have as elements the subband-dependent functions 
$G^{mm'}_{{\bf k}\sigma}(E)$, $T^{mm'}_{{\bf k}}$, and 
$M^{mm'}_{{\bf k}\sigma}(E)$, respectively.

The evaluation of the higher Green functions resulting from
(\ref{eom_gen}) employs a moment-conserving decoupling approximation
(MCDA). Following the steps presented in \cite{SN99b} one arrives at an
implicite set of equations for the local self-energy, 
$M^{mm'}_{{\bf k}\sigma} (E) \equiv \delta_{mm'} M^m_{\sigma} (E)$:
\begin{equation}
  \label{small_self}
  M^m_{\sigma}(E) = -\frac{J}{2} \mu^m_{\sigma} (E), \quad
  \mu^m_{\sigma} (E) =
  \frac{Z^m_{\sigma}(E)}{N^m_{\sigma}(E)},
\end{equation}
where the numerator and the denominator, respectively, are given by
\begin{eqnarray}
  \label{num_den_z}
  Z^m_{\sigma} & = & z_{\sigma} \hslash^2 \langle S^z
  \rangle + \frac{J}{2} \Big\{ \big( \kappa^{(2)}_{\sigma} -
  \hslash^2 S(S+1) \big) G^{mm}_{{\bf 0}\sigma}
  \nonumber \\ & &
  - \big( (\lambda^{(1)}_{\sigma} +
  \lambda^{(2)}_{\sigma} + z_{\sigma} \mu^m_{-\sigma} )
  \hslash \langle S^z \rangle + \hslash
  \kappa^{(2)}_{\sigma} \big) G^{mm}_{{\bf 0}-\sigma}
  \Big\}
  \nonumber \\ & & + \frac{J^2}{4} \Big\{
  z_{\sigma} \hslash^2 S(S+1)
  (\lambda^{(1)}_{\sigma} + \lambda^{(2)}_{\sigma}
  + z_{\sigma} \mu^m_{-\sigma}) 
  \nonumber \\ & & 
  + \kappa^{(2)}_{\sigma} (\mu^m_{\sigma} - \mu^m_{-\sigma}) \Big\} 
  G^{mm}_{{\bf 0}\sigma} G^{mm}_{{\bf 0}-\sigma} ,\\
  \label{num_den_n} N^m_{\sigma} & = & \hslash^2 - \frac{J}{2} \Big\{
  (\hslash + z_{\sigma} \lambda^{(2)}_{\sigma} + \mu^m_{\sigma})
  G^{mm}_{{\bf 0}\sigma} + (z_{\sigma} \lambda^{(1)}_{\sigma} 
  \nonumber \\ & &
  + \mu^m_{-\sigma}) G^{mm}_{{\bf 0}-\sigma} \Big\} + \frac{J^2}{4}
  \Big\{ (\mu^m_{\sigma} + \hslash) (\mu^m_{-\sigma} + z_{\sigma}
  \lambda^{(1)}_{\sigma}) \nonumber \\ & &
  + z_{\sigma} \lambda^{(2)}_{\sigma} (\mu^m_{-\sigma} + \hslash) \Big\}
  G^{mm}_{{\bf 0}\sigma} G^{mm}_{{\bf 0}\sigma},
\end{eqnarray}
where
\begin{eqnarray}
  \label{coeff} 
  & \kappa^{(1)}_{\sigma}=0, \quad \kappa^{(2)}_{\sigma} =
  \langle S^{-\sigma} S^{\sigma} \rangle
  - \lambda^{(2)}_{\sigma} \langle S^z \rangle, & 
  \nonumber \\
  & \lambda^{(1)}_{\sigma} = 
    \displaystyle \frac{\langle S^{-\sigma}
    S^{\sigma} S^z \rangle + z_{\sigma}
    \langle S^{-\sigma} S^{\sigma} \rangle}
    {\langle S^{-\sigma} S^{\sigma} \rangle}, &
  \\
  & \lambda^{(2)}_{\sigma} =
    \displaystyle \frac{\langle S^{-\sigma}
    S^{\sigma} S^z \rangle - \langle S^z
    \rangle \langle S^{-\sigma} S^{\sigma} \rangle}
    {\langle (S^z)^2 \rangle - \langle S^z \rangle^2}.
    & \nonumber
\end{eqnarray}
contain {\em f}-spin correlation functions, which have to
be calculated within the Heisenberg model. 

For the limiting case of ferromagnetic saturation of the localized 
{\em f}-spin system ($T=0$), there exists an {\em exact} solution for
the {\em d-f} ({\em s-f}) model with empty condcution band ($n=0$)
\cite{SMN97}. In particular, the spectrum of the spin-$\uparrow$
electron is only rigidly shifted compared to the free solution obtained
for the case of vanishing {\em s-f} interaction, $J=0$. This means, that
the spin-$\uparrow$ spectra obtained within a ferromagnetic
band-structure calculation provide the single-particle input
(cf.~Eq.~(\ref{h_d})) for the {\em d-f} model calculation.
Since the solution of the {\em d-f} for $T=0$ and $n=0$ is exact, 
the problem of double counting of relevant interactions is elegantly
avoided in the presented approach. 

To obtain the temperature-dependent band structure of EuO,
the $T=0$ band structures are needed as single-particle input for the
model calculations. These have been calculated for the 
Eu-{\em 5d} bands, which in EuO dominate the conduction band, using a
standard TB-LMTO program \cite{And75}. The results of these calculations
will be presented in the next section.

\section{Band-structure calculations}
The europium chalcogenides crystallize in the fcc rock-salt
structure. For EuO, the lattice
constant is $a=5.142\,{\rm \AA}=9.717\,{\rm au}$ \cite{HMGM70,Wac91}.
The electronic configurations of europium and oxygen in the state of the
free atom and in the ionic picture for ${\rm EuO}$, respectively, are
\begin{displaymath}
  \begin{array}{l@{\quad}l@{\quad}l}
    {\rm free\!:} & [{\rm Eu}] = [{\rm Xe}](4f)^7 (6s)^2, &
    [{\rm O}] = [{\rm He}] (2s)^2 (2p)^2; \\[1ex] {\rm EuO:} &
    [{\rm Eu}] = [{\rm Xe}](4f)^7, & [{\rm O}] = [{\rm Ne}].
  \end{array}
\end{displaymath}
In accordance with Hund's coupling, the ${\rm Eu}$-{\em 4f}\/ electrons
couple to the 
maximum magnetic moment of $S=7/2$. The difficulty in dealing
with the {\em 4f}\/ levels within an LDA calculation lies in their strongly
localized character, which is due to a strong Coulomb interaction
between the {\em 4f}\/ electrons. As a result of the inability of the
LDA to take into account the respective interactions correctly, a 
{\em normal}\/ LDA calculation for ${\rm EuO}$ produces a metal with the
{\em 4f}\/ levels lying well within the conduction band. 
 
\begin{figure}[t!]
  \centerline{\includegraphics[width=\linewidth]{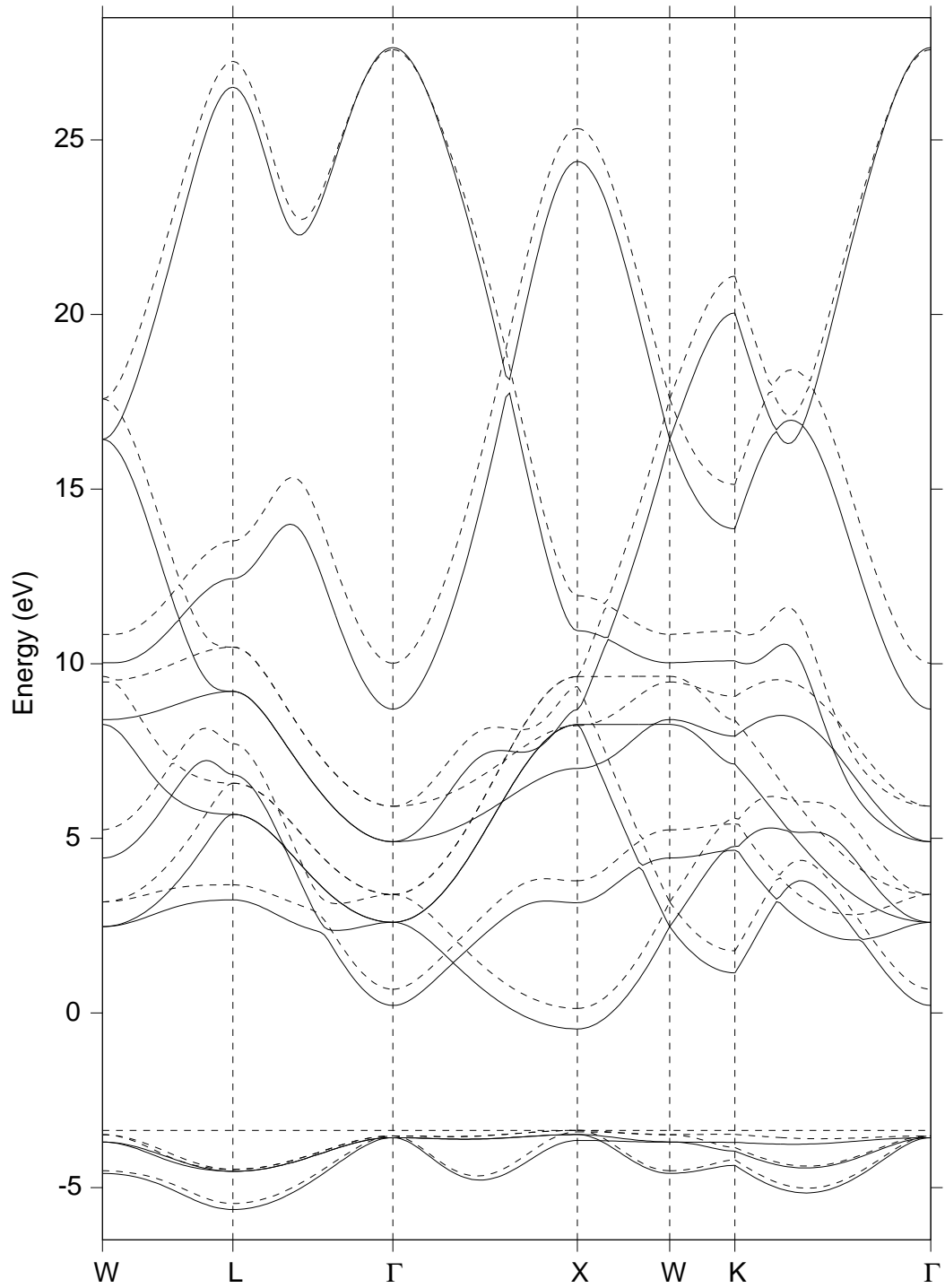}}
  \caption{Spin-dependent (spin-$\uparrow$: ---$\!$---, spin-$\downarrow$:
    - - -) band structure of bulk ${\rm EuO}$ calculated
    within an LSDA calculation with the {\em 4f}\/  levels treated as core
    electrons. The horizontal dashed line represents the Fermi energy.} 
  \label{fig1}
\end{figure}

\begin{figure}[t!]
  \centerline{\includegraphics[width=\linewidth]{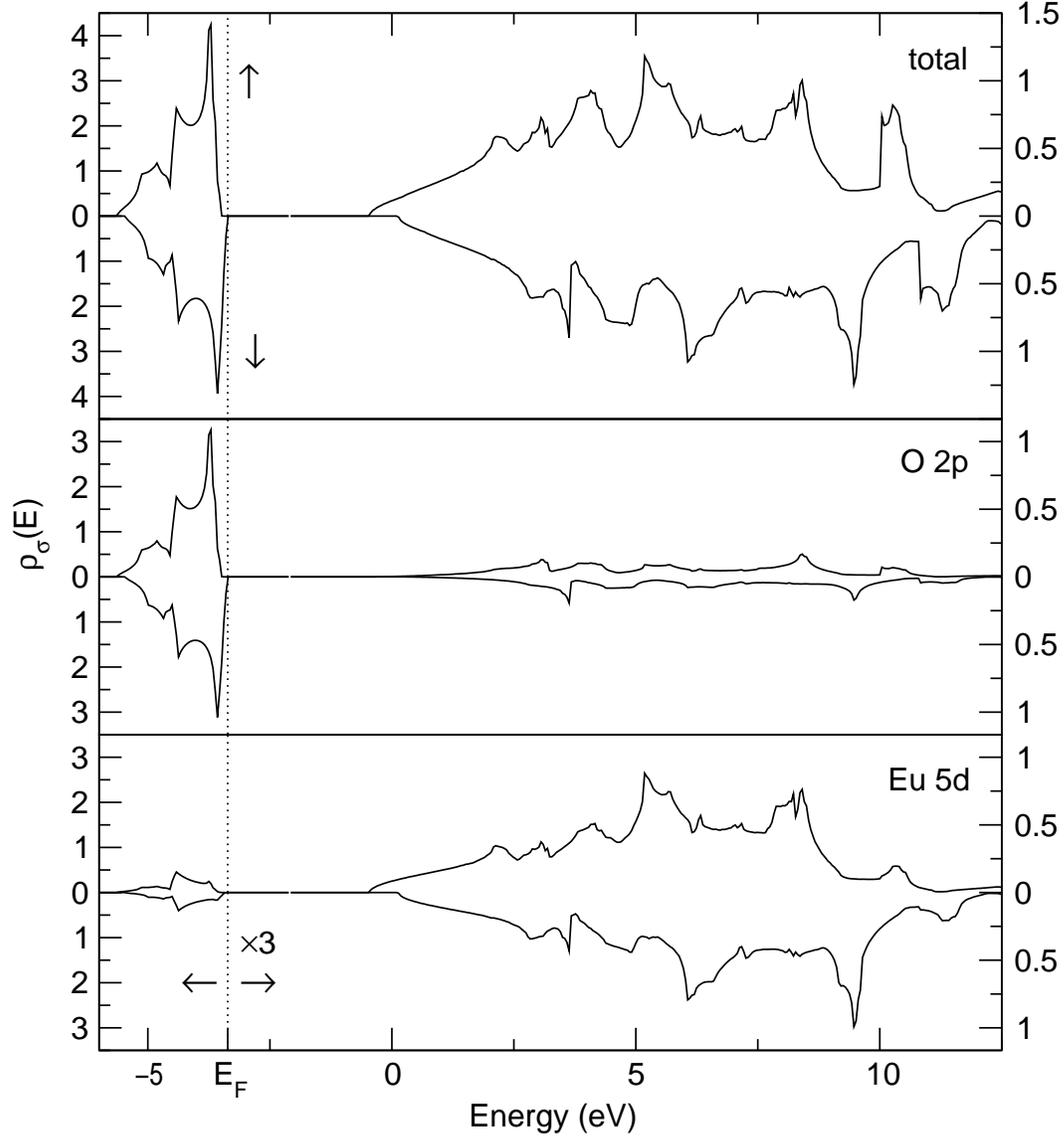}}
  \caption{Density of states, $\rho_{\sigma}(E)$, of bulk ${\rm EuO}$
    calculated within an LSDA calculation with the {\em 4f}\/ levels
    treated as core electrons. The upper graph shows the total density
    of states, whereas the lower two display the partial density of
    states of the ${\rm O}$-{\em 2p}\/ and the ${\rm Eu}$-{\em 5d}\/
    bands. The dotted line represents the Fermi energy. The densities of
    states above the Fermi energy are multiplied by a factor of three
    and correspond to the left $y$-axis.} 
  \label{fig2}
\end{figure}

To overcome this situation, we chose to treat the {\em 4f}\/ moments as
localized core electrons\footnote{For EuO, it has been shown \cite{Schiller00}
  that the band structures of the relevant conduction bands obtained
  within such a calculation differ only slightly from those obtained
  within an LDA+U caculation. For the {\em 4f}\/ system ${\rm Gd}$ the 
  different approaches for dealing with the localized {\em 4f}\/ moments
  have been reviewed by Eriksson {\em et al.}\cite{EAO+95}.}. 
Fig.~\ref{fig1} shows the respective spin-dependent
band structure of ${\rm EuO}$. In Fig.~\ref{fig2}, for
the same calculation, the density of states is displayed. Here, the main
contributions to the density of states originate from the 
${\rm O}$-{\em 2p}\/ levels, which constitute the valence band, and from
the unoccupied ${\rm Eu}$-{\em 5d}\/ levels. As a result of treating the
{\em 4f}\/ levels as core electrons, we have the correct ground state of
an insulator.  Clearly, the conduction-band region is dominated by the 
${\rm Eu}$-{\em 5d}\/ electrons. Thus, for the model calculations which
will be performed in Sec.~\ref{sec:temp-depend-bandstr} it is a
reasonable approximation to restrict the single-particle input obtained
within the band-structure calculations to the ${\rm Eu}$-{\em 5d}\/ bands. 

\section{Temperature-dependent band structures}
\label{sec:temp-depend-bandstr}
In this section the LSDA band structure of bulk ${\rm EuO}$ shall be
combined with an {\em d-f}\/ model calculation to obtain the
temperature-dependent band structure. Due to the fact that within our model
calculation for $T=0$ the spin-$\uparrow$ spectrum is rigidly shifted
towards lower energies by the amount of $\frac{1}{2}JS$, the obtained
{\bf k}-dependent hopping matrices for the spin-$\uparrow$ electron will
serve us as the input for the kinetic Hamiltonian~(\ref{h_d}). Since
the solution for $T=0$ is exact, the problem of double counting of
relevant interactions, which usually occurs when combining
first-principles and model calculations, is elegantly avoided.

The temperature comes into play via the
temperature-dependent {\em f}-spin correlation functions of
Eqs.~(\ref{coeff}), which can be calculated within the 
Heisenberg model for the local-moment system
\cite{Schiller00,SN99b,SN99a}. 
For the following investigations, however, the concrete shape of the
magnetization curve is of lesser interest and the reduced 
magnetization $\langle S_z \rangle / S$ has been chosen as the
temperature parameter. 

To calculate the temperature-dependent band structure of EuO, the 
{\em d-f} exchange interaction is needed.
It has been shown by several authors\cite{JW76,PKA76,Gun77}, that the
standard LSDA band-structure calculations are quite compatible with the
simple Stoner or mean-field picture, in which the exchange splitting is
only slightly energy dependent. 
Contrary to this statement, Fig.\ref{fig2} implies an
energy-dependent splitting of the spectra.
However, the inclusion of a such an energy-dependent splitting
of the spectra is not possible within the theory presented in
this paper and we have to take an averaged exchange
interaction for the calculation of the temperature-dependent band
structures. Averaging the splitting of the spectra for
the two spin directions over the energy, one arrives at an average
splitting of the spectra for the two spin directions of 
$0.875\,{\rm eV}$. In the mean-field approximation of the {\em d-f}\/
model for $T=0$ the spin-splitting is equivalent to $JS$. With $S=7/2$
one gets for the {\em d-f}\/ exchange 
splitting $J=0.25\,{\rm eV}$. This value will be used for the
further {\em d-f}\/ model calculations.

\begin{figure}[t!]
  \centerline{\includegraphics[width=\linewidth]{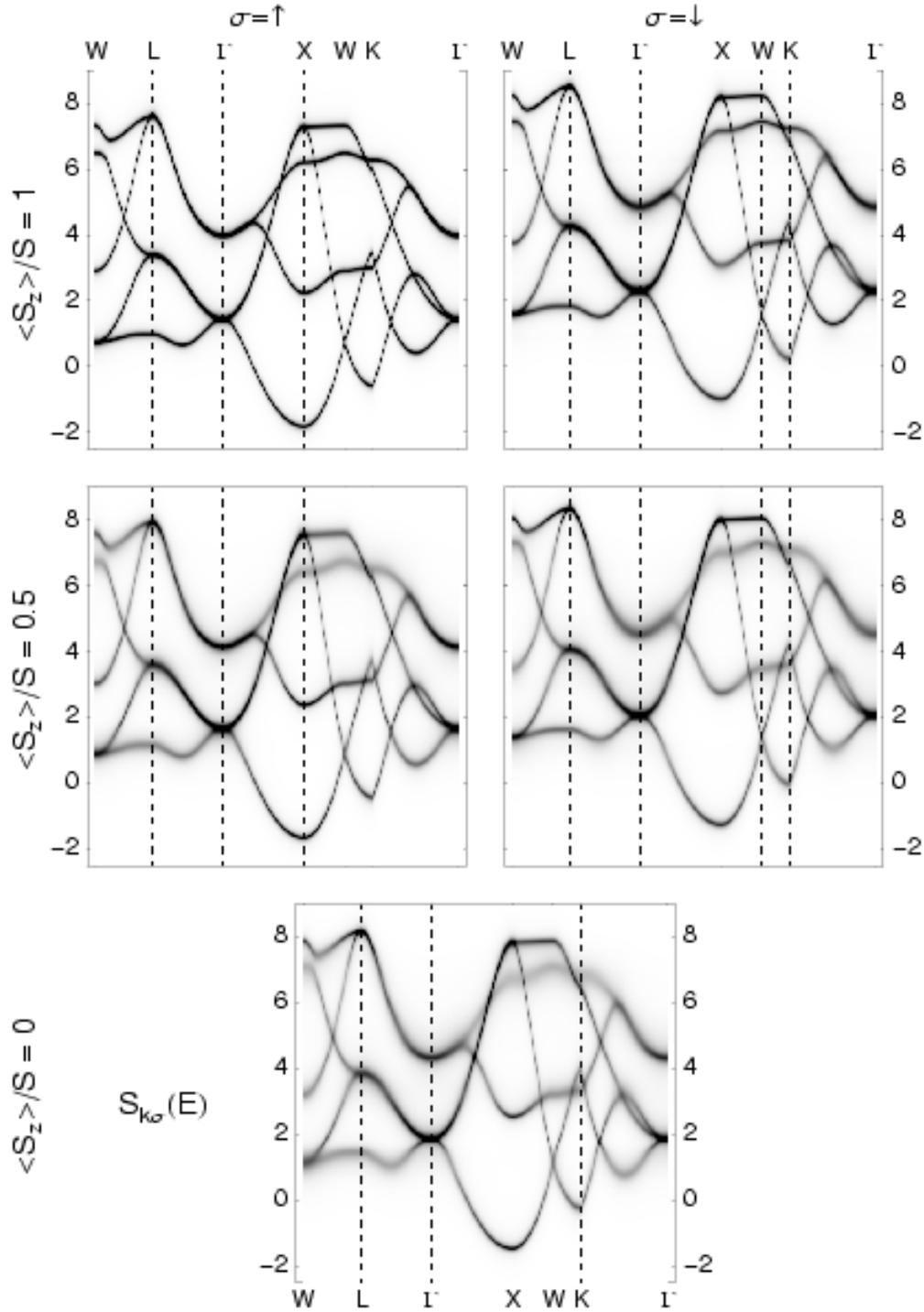}}
  \caption{Spin-dependent spectral densities of the ${\rm Eu}$-{\em 5d}\/
    bands of bulk ${\rm EuO}$ for \mbox{$J=0.25\,{\rm eV}$} 
    and for different magnetizations $\langle S_z \rangle/S$.}   
  \label{fig3}
\end{figure}

\begin{figure}[t!]
  \centerline{\includegraphics[width=\linewidth]{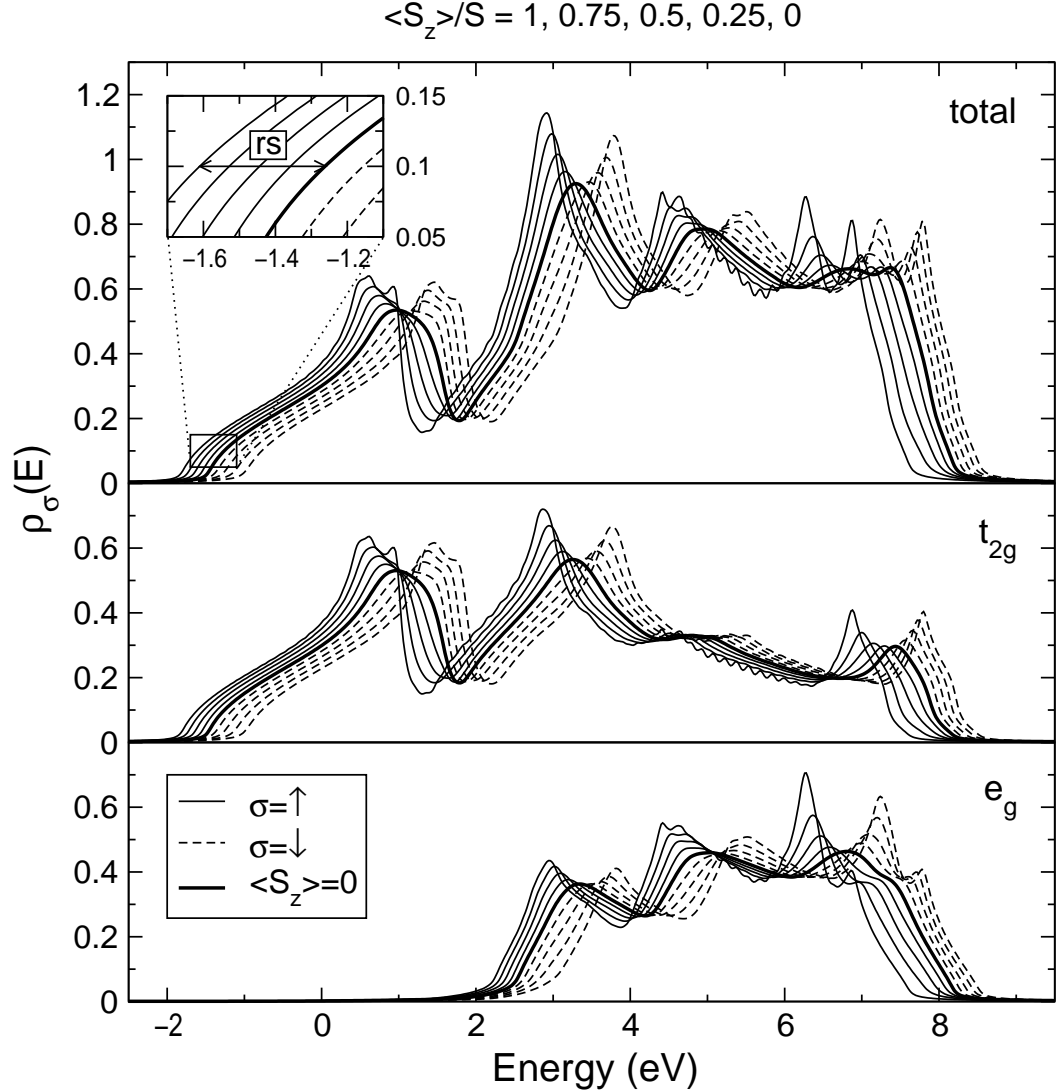}}
  \caption{Temperature-dependent densities of states of the 
    ${\rm Eu}$-{\em 5d}\/ bands of bulk ${\rm EuO}$. For $T=0$ 
    ($\langle S_z \rangle/S=1$) 
    the spectra for the two spin directions are furthest
    away from each other and approach each other when increasing the
    temperature. For $T=T_{\rm C}$ ($\langle S_z \rangle /S =0$) the densities
    of states for both spin directions are the same ({\bf fat} lines).}  
  \label{fig4}
\end{figure}

Fig.~\ref{fig3} shows the temperature-dependent spectral density
of bulk ${\rm EuO}$ obtained within an {\em d-f}\/ model calculation with
$J=0.25\,{\rm eV}$. For $T=0$ ($\langle S_z \rangle /S =1$), the
spin-$\uparrow$ spectral density agrees, except for a constant energy shift,
with that obtained by the LSDA calculation. However, still for $T=0$ but
for the spin-$\downarrow$ spectrum a broadening of the dispersion curves
can be observed, most notably around the $\Gamma$-point, which indicates a
finite lifetime of the respective 
quasiparticles due to correlation effects. In this respect already the
$T=0$ spin-$\downarrow$ solution goes beyond LSDA by taking into account
correlation more realistically.

For intermediate temperatures ($\langle S_z \rangle / S=0.5$) a
broadening of the dispersion curves  sets in also for the
spin-$\uparrow$ spectra 
since the spin-$\uparrow$ electron can now exchange its spin with the
deviated local-moment system. For the spin-$\downarrow$ spectral
density, the correlation effects, already present at $T=0$, increase, as
can be seen by the further broadening of the curves. At the same time,
the spin-$\uparrow$ and the spin-$\downarrow$ spectra are shifted
towards higher and lower energies, respectively, reducing the effective
splitting between the two spectra. 
This effect continues with increasing temperature, until finally, for
$T=T_{\rm C}$ ($\langle S_z \rangle /S=0$), the spectra for both spin
directions are equal. Clearly, the temperature-dependent effects
displayed in Fig.~\ref{fig3} do not comply with the simple
Stoner-picture which would imply a constant energy shift of the
spectra. The reason for this more complicated behavior as a function of
temperature is again that in the theory, the correlation is treated in a
way which goes beyond mean-field. 

The mentioned shift of the spin-$\uparrow$ spectrum towards lower
energies when going from $T=T_{\rm C}$ down to $T=0$ represents the red
shift of the optical absorption edge in ${\rm EuO}$\cite{Wac64,BJW64}.
Fig.~\ref{fig4} displays the densities of states obtained for
different magnetizations of the {\em 4f}\/ moments. It can be seen
that the red shift of the optical absorption edge is due to the 
{\em 4f}-${\it 5d}_{t_{2g}}$ transition. From Fig.~\ref{fig4} one
obtains a red shift of ${\rm rs}=0.35\,{\rm eV}$. This value
agrees reasonably with the experimental value for the red shift of
$0.27\,{\rm eV}$ \cite{Wac91}. Here, the agreement can 
be further improved when choosing the splitting of the lower edges of
the LSDA ${\rm Eu}$-{\em 5d}\/ bands in Fig.~\ref{fig2} to calculate
the {\em d-f}\/ exchange interaction for the model calculation. 

There have been previous calculations by Nolting {\em et al.}\  concerning
the temperature-dependent band structure of bulk ${\rm EuO}$
\cite{NBB87a,NBB87b}.  
In these works, the band structure of the ${\rm Eu}$-{\em 5d}\/ bands has
been split into five {\em s}-like bands with the lowest eigenstates
belonging to the lowest band etc. The splitting into {\em s}-bands
produces five consecutive bands of an average bandwidth of about
$W=3\,{\rm eV}$\cite{NBB87a,NBB87b}.  

Contrary to these works, for the calculation of
Figs.~\ref{fig3} and~\ref{fig4} the full
band structure of the ${\rm Eu}$-{\em 5d}\/ bands has been taken into
account, thereby respecting the symmetry of the different 
${\rm Eu}$-{\em 5d}\/ orbitals. As a result, the ${\rm Eu}$-{\em 5d}\/ bands
of bulk ${\rm EuO}$ split into the $t_{2g}$ and the $e_g$ subbands
with a bandwidth of about $10\,{\rm eV}$ and $6\,{\rm eV}$, respectively.
For the model calculations, the decisive entity for the magnitude of the
correlation effects is the {\em d-f}\/  exchange interaction over
bandwidth, $J/W$.  
In this respect, in the previous works \cite{NBB87a,NBB87b} the
calculated correlation effects should be slightly overestimated,
considering the more ``natural'' band decomposition employed in this
thesis. 

\section{Conclusion}
We presented a method for calculating the temperature-dependent
band-structure of a local-moment system based on a parametrization of
the (multi-band) {\em d-f} model Hamiltonian using results of
first-principles TB-LMTO band structure calculations. The method has
been applied to the ferromagnetic semiconductor EuO.

The presented method enables us to calculate the temperature-dependent
bandstructure of EuO over the entire temperature range. A main
characteristics of the approach is the avoidance of double-counting of
relevant interactions for the conbination of first-principles and
model-calculation.

Concerning the earlier works on the temperature-dependent band structure
of bulk EuO \cite{NBB87a,NBB87b}, the method presented in this paper has
two main advantages:
\begin{enumerate}
\item The full multi-band character of the conduction bands, including
  in particular the symmetry of the relevant orbitals, is taken into
  account. 
\item Due to the broader bandwidths of the involved subbands the
  correlation effects should be treated in a more realistic fashion.
\end{enumerate}
The extension of the method to systems with finite band occupation,
$n\ne 0$, and to systems with reduced dimensionality promises  to give
an insight into the fascinating physics at the surface of rare-earth
systems \cite{DMIL97}.

\section*{Acknowledgements}
This work was supported by the German Research Foundation (DFG) within
the Sonderforschungsbereich 290. One of the authors (R.\,S.) would like
to acknowledge the support by the German National Merit Foundation.


\end{document}